\documentclass[conference]{IEEEtran}
\usepackage{graphicx}
\usepackage[hyphens]{url}
\usepackage{hyperref}
\hypersetup{breaklinks=true}

\usepackage{listings}
\usepackage{xcolor}
\usepackage{amsmath}
\usepackage{subcaption}
\usepackage{pgfplots}

\newcommand{\eg}{\textit{e.g.,~}}

\newcommand{\ie}{\textit{i.e.,~}}
\newcommand{\etal}{\textit{et al.}}

\definecolor{backcolor}{RGB}{173, 239, 209}
\definecolor{textcolor}{RGB}{0, 32, 63}
\definecolor{codegreen}{rgb}{0,0.6,0}
\definecolor{backcolour}{rgb}{0.95,0.95,0.92}
\definecolor{codegray}{rgb}{0.5,0.5,0.5}
\definecolor{myOrange}{RGB}{232, 112, 0}
\definecolor{myTeal}{RGB}{2, 136, 143}
\definecolor{myGreen}{RGB}{28, 53, 45}
\definecolor{myGray}{RGB}{240, 240, 240}



\makeatletter
\newcommand{\thickhline}{%
    \noalign {\ifnum 0=`}\fi \hrule height 1pt
    \futurelet \reserved@a \@xhline
}

\begin{document}

\title{ss2DNS: A Secure DNS Scheme in Stage~2}

\author{\IEEEauthorblockN{Ali Sadeghi Jahromi}
	\IEEEauthorblockA{Carleton University\\
		alisadeghijahromi@cmail.carleton.ca}
	\and
	\IEEEauthorblockN{AbdelRahman Abdou}
	\IEEEauthorblockA{Carleton University\\
		abdou@scs.carleton.ca}
	\and
	\IEEEauthorblockN{Paul van Oorschot}
	\IEEEauthorblockA{Carleton University\\
		paulv@scs.carleton.ca}}

\maketitle

\begin{abstract}
The absence of security and privacy measures between DNS recursive resolvers and authoritative nameservers has been exploited by both on-path and off-path attackers. Although numerous security proposals have been introduced in practice and in the literature, they often face deployability barriers and/or lack a compelling set of security and privacy properties, resulting in limited adoption. We introduce ss2DNS, a novel DNS scheme designed to mitigate the security and privacy vulnerabilities in the resolution process between resolvers and authoritative nameservers, while preserving efficiency by maintaining a single round-trip. ss2DNS takes advantage of a hierarchical trust model that does not rely on entities external to DNS zones, and \textit{delegates} nameserver replicas within each zone to serve zone data securely for short, renewable time intervals. This design enables real-time security properties for DNS messages without requiring the duplication of long-term private keys on replicas, thereby minimizing exposure to compromise. We implement a proof of concept of ss2DNS for evaluation and show that for server-side processing latency, resolution time, and CPU usage, ss2DNS is comparable to less-secure schemes but significantly outperforms DNS-over-TLS.
\end{abstract}

\IEEEpeerreviewmaketitle

\section{Introduction}
\label{sec:intro}
\begin{figure}
    \centering
    \includegraphics[width=\linewidth]{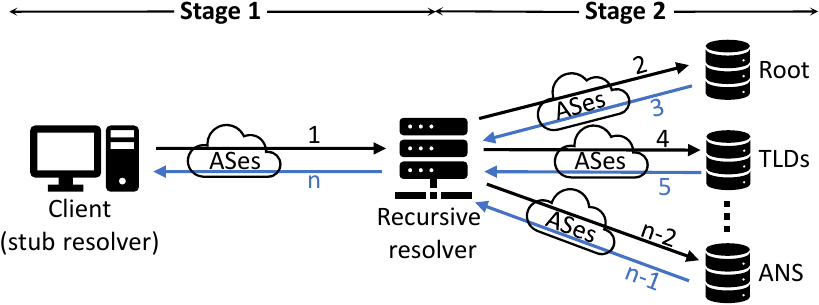}
    \caption{Two stages of DNS resolution process.}
    \label{fig:nameRes}
\end{figure}

The Domain Name System (DNS) was originally introduced to map domain names to their corresponding IP addresses~\cite{rfc1034, rfc1035}. As shown in Fig.~\ref{fig:nameRes}, DNS typically operates as a two-stage protocol. Stage~1 involves communication between a stub resolver and a recursive resolver, while Stage~2 consists of interactions between the recursive resolver and authoritative name servers (ANSes). In Fig.~\ref{fig:nameRes}, the client uses a public recursive resolver over the Internet. As a result, one or more Autonomous Systems (ASes), including the client’s Internet Service Provider (ISP), may exist between the client and the recursive resolver. Similarly, different ASes may exist between the recursive resolver and the ANSes of different zones in Stage~2. DNS forwarders are also common in Stage~1, but omitted from the figure for simplicity. While the goals of efficiency, availability, and scalability were effectively met in what we refer to as \textit{Vanilla DNS}~\cite{rfc1034, rfc1035} (the original design), security and privacy were not part of the initial design goals. The absence of security in DNS introduces vulnerabilities across various applications that depend on it, potentially leading to the issuance of fraudulent certificates~\cite{birge2018bamboozling, brandt2018DV++}, time synchronization failures, and privacy and availability risks for services such as email and VPNs~\cite{DaiDNS21}. Moreover, the lack of privacy in DNS facilitates inference of application usage and device types~\cite{apthorpe2017closing}, as well as large-scale surveillance of DNS resolution activity~\cite{morecow2017nsa, NSASpy}.

Over the years, numerous name resolution schemes have been proposed to enhance the security and privacy of the DNS resolution process, thereby mitigating the vulnerabilities of vanilla DNS~\cite{zhu2015TDNS, DoH-rfc8484, rfc4033, curvedraft}. Proposals that involve fundamental changes to the DNS infrastructure~\cite{kalodner2015namecoin, rfc9498GNS} generally face significant adoption barriers. We believe that a more practical approach is to rely on existing DNS infrastructure and its trust model to increase the chance of adoption of improved proposals.

The majority of secure DNS proposals focus on Stage~1 of the name resolution process\cite{DNSCryptV2, DoH-rfc8484, zhu2015TDNS, doq-rfc9250}. Stage~1 schemes, such as DNS-over-TLS (DoT)\cite{DoT-rfc7858} and DNS-over-HTTPS (DoH)~\cite{DoH-rfc8484}, have received significant attention and have been adopted by major client-side web browsers and public recursive resolvers.\footnote{\url{https://developers.cloudflare.com/1.1.1.1/encryption/dns-over-https/encrypted-dns-browsers/}} In contrast, Stage~2 secure DNS schemes have not achieved widespread adoption~\cite{rfc4033, curvedraft, duan2023rhine, confdns}. The Domain Name System Security Extensions (DNSSEC) is the only Stage~2 scheme that has been widely adopted by the root and Top-Level Domains (TLDs), although its deployment among subordinate zones remains limited (approximately 7\% in Second-Level Domains (SLDs)\cite{yajima2021measuring})\footnote{\url{https://rick.eng.br/dnssecstat/}}, with adoption gradually increasing~\cite{chung2017longitudinal, yajima2021measuring}. The primary reasons for this include the absence of a compelling set of security or privacy properties, as well as deployability-related challenges~\cite{chung2017longitudinal, cowperthwaite2010futility, IanaDNS}.

Given the differing requirements and proposed schemes in each stage, securing the DNS resolution path can be considered as two distinct stages. Accordingly, Stage~1 and Stage~2 schemes are compatible and can be deployed in parallel to enhance the overall security of the DNS resolution process.

Due to the lack of a widely adopted secure DNS scheme in Stage~2, both on-path and off-path adversaries remain capable of injecting false DNS responses and poisoning the caches of recursive resolvers. Such cache poisoning attacks in Stage~2 are particularly severe, as the injected false responses are stored at the resolver and subsequently served to significantly large number of clients querying the resolver for the affected domain. In contrast, in Stage~1, an injected DNS response typically poisons only the local cache of a single client, thereby limiting the scope of the attack. To address the existing gap in Stage~2, we propose ss2DNS, a secure Stage~2 DNS scheme that enhances name resolution security and privacy with minimal impact on performance and moderate computational overhead; we do so through insights drawn from the strengths and limitations of previous secure DNS schemes.

To achieve these security and privacy benefits, we introduce a novel short-term delegation mechanism that enables a DNS zone to delegate the signing of cryptographic parameters to its nameserver replicas (entities that may not be fully trusted by the zone owner). Short-term delegation approaches, such as the Delegated Credentials mechanism proposed in RFC 9345~\cite{DCrfc9345} for Content Delivery Networks (CDNs) in TLS-based communications, have demonstrated significant advantages in mitigating key compromise risks. This short-lived delegation enables nameservers to perform real-time cryptographic operations, while reducing the exposure of long-term cryptographic keys and minimizing the risks associated with short-term key compromise through automatic key revocation~\cite{AbdouRevocation20}. Furthermore, this short-term delegation mechanism eliminates the deployability barrier that hindered the adoption of DNSCurve at the root zone due to the need for long-term secret replication~\cite{IanaDNS}. Moreover, ss2DNS employs a reverse-tree hierarchical trust model already supported by root and TLDs, requiring no additional infrastructure.

To evaluate the performance of ss2DNS, we implement a Proof of Concept (PoC) and compare three metrics: server-side processing latency, resolution time, and CPU utilization. We evaluate ss2DNS compared to other schemes, including Vanilla DNS~\cite{rfc1034}, DNSSEC~\cite{liveDNSSEC}, and DoT~\cite{DoT-rfc7858}. Preliminary results indicate that ss2DNS offers performance comparable to less secure schemes, while significantly reducing resolution delay and CPU usage relative to DoT.

The key contributions of our work are as follows:
\begin{itemize}
    \item We introduce a novel short-term delegation mechanism within DNS zones to avoid replicating long-term keys, reduce key compromise risks, and facilitate deployment.
    \item We design ss2DNS, a novel secure DNS scheme that enhances the security and privacy of name resolution in Stage~2. ss2DNS encapsulates DNS queries and responses without modifying the original DNS zones or messages, or adding extra network round-trips.
    \item We implement and evaluate a PoC, analyzing the impact of ss2DNS on CPU utilization, resolution time, server-side processing time, and fragmented messages. We will make the implementation publicly available.
\end{itemize}

\section{Background: DNSSEC and DNSCurve}
\label{sec:background}

Multiple DNS schemes have been proposed with the goal of improving the security and privacy of DNS resolution in Stage~2. We provide a background on two prominent schemes: DNSSEC~\cite{rfc4033, rfc4034, rfc4035} and DNSCurve~\cite{curvedraft, DNSCurveweb}.

\subsection{DNSSEC}
\label{sssec:DNSSECFlow}
To mitigate false response injections in Stage~2, DNSSEC~\cite{rfc2065} was introduced in the 1990s to ensure DNS response integrity and authenticity. It has since been revised to its current standard, as defined in RFCs 4033-4035~\cite{rfc4033, rfc4034, rfc4035}. DNSSEC was designed to augment Vanilla DNS by adding authenticated denial of existence, message authentication, and integrity. DNSSEC is typically implemented in Stage~2, leaving stub-to-recursive communication unsecured. In Stage~1, a DNSSEC-enabled recursive resolver can inform clients via the Authenticated Data (AD) header bit that a DNS response has been successfully authenticated using DNSSEC. However, Stage~1 needs to be secured separately, \eg through DoH~\cite{DoH-rfc8484}. Also, the client must either trust the recursive resolver or do the resolution itself.

\textbf{Message Authentication in DNSSEC}: DNSSEC introduces two asymmetric keys as DNSKEY records in each zone: Zone Signing Key (ZSK) and Key Signing Key (KSK). The KSK is used exclusively to sign DNSKEY records within the zone. Based on the local policy within a zone, KSK can be considered as the long-term zone key and renewed less frequently~\cite{rfc4033}. On the other hand, ZSK within a zone is used for signing all the resource records, except for the DNSKEY records, for which the zone is authoritative. The ZSK can be defined with a shorter lifetime and renewed more often than the KSK~\cite{rfc4033}. Whenever a resolver queries a DNS record from an ANS within a DNSSEC-protected zone, the digital signature of that record, known as Resource Record Signature (RRSIG), is also included in the DNS response that the ANS returns. Subsequently, in order to authenticate the RRSIG contained in DNSSEC responses, the resolver sends an additional query to an ANS of the zone to obtain the DNSKEY records (\ie ZSK and KSK) of the zone. At this point, the resolver verifies that the original DNS record is signed by the ZSK of the zone. Additionally, the resolver verifies that the DNSKEY record containing the ZSK is signed by the KSK of the zone. Thus, the resolver can authenticate the queried standard DNS record using the zone keys.

\textbf{Trust Model in DNSSEC}: In addition to verifying the authenticity of RRSIGs of DNS records, a DNSSEC-enabled resolver requires a means to trust that a KSK belongs to a specific zone. Thereby, the resolver can trust the ZSK's RRSIG, which is generated by the zone's KSK. With this, the ZSK can be trusted, which can be used to trust RRSIGs of other resource records, which are generated by the ZSK. Lastly, the desired resource records can be trusted.

So as to form the trust model of DNSSEC, the public KSK of the root zone is defined as the trust anchor of DNSSEC and included in the recursive resolver software. TLD zones, which are delegated from the root, send the hash of their KSK DNSKEY records, known as the Delegation Signer (DS) record, to the root zone. Similarly, subordinate zones beneath the TLDs send their DS records to their respective parent zones. These DS records, received from subordinate zones, are signed by the ZSK of their parent zone. Therefore, the collection of signed DS records of child zones within their parent zone establishes a chain of trust extending from each zone to the root. A resolver can start from the root zone (trust anchor) and, by using the DS records of TLDs in the root, validate the KSKs of the TLDs. It can then continue down the hierarchy, using each zone’s DS record in its parent to validate the KSK of the child zone.

After establishing the chain of trust, recursive resolvers trust the KSK of the root zone, and thereby they can validate the signature of the ZSK of the root zone and thus trust other records within the root, including the DS records of the TLDs. Since ZSK of the root zone signs the DS records of its child zones (TLDs), a resolver can obtain the DS records of the TLD directly from the root. Subsequently, during interaction with a TLD server, the resolver verifies that the hash of the TLD's KSK matches the signed DS record received from the root zone. As the resolver traverses the DNS hierarchy, it can authenticate the KSKs of zones by checking the signed DS record within their parent zone, thereby establishing trust in their KSKs, ZSKs, and RRSIGs within DNSSEC-protected zones and authenticating standard DNS records.

\textbf{Caching DNSKEYs}: Like other DNS records, DNSKEYs have a Time To Live (TTL) field (a 32-bit value indicating how long resolvers should cache them). Short TTLs offer more flexibility in responding to key compromise but increase computational and bandwidth load on ANSes and resolvers. They also lead to longer resolution times, as expired keys require resolvers to query these keys for authentication. On the other hand, long TTLs for the zone keys result in a lack of flexibility in the key compromise situations. However, larger TTL values for DNSKEYs improve the name resolution performance, as the keys are queried less frequently as their presence in the resolver caches is more likely. Taking both sides into account, the caching time should neither be excessively long to mitigate the damage of key compromise situations nor very short to minimize the name resolution delay.

\textbf{Reflection Amplification}: DNSSEC uses UDP as transport layer protocol with Extension mechanisms for DNS (EDNS(0))~\cite{rfc6891}, which enables transmitting DNS responses larger than the original DNS maximum response (512 bytes) over UDP. These design choices, alongside the added signatures and keys in DNSSEC responses, enable reflection amplification attacks with significant amplification factors up to $100\times$~\cite{DNSCurveweb, DNSSECAmp}. Thus, attackers can exploit DNSSEC to amplify the traffic of their Distributed Denial of Service (DDoS) attacks by sending queries that produce larger responses directed at targeted servers~\cite{cowperthwaite2010futility, DNSSECAmp}.

\textbf{Unsigned Records}: In DNSSEC, non-authoritative delegating records within DNSSEC-protected zones are not signed. Specifically, glue and NS resource records of child zones are not part of the authoritative DNS data secured in the parent zones. Thus, these non-authoritative records in the parent zones are transferred unsigned~\cite{rfc4033, herzberg2013towards}. Therefore, these unsigned records do not benefit from the security properties of DNSSEC. The injection and caching of these unsigned records in a validating recursive resolver can result in DNSSEC validation failure, potentially causing disruptions in the resolution of DNS queries when attempting to access the legitimate nameservers. Additionally, in instances where the recursive resolver falls back to Vanilla DNS or accepts unauthenticated responses, these unsigned records can result in the injection of false responses and \textit{downgrade attacks}~\cite{herzberg2013towards}.

\textbf{Zone Enumeration}: A DNSSEC-enabled zone requires a specific type of signed resource record to indicate that a record does not exist within a zone. Initially, DNSSEC used Next Secure (NSEC) records to provide authenticated denial of existence for a non-existent DNS record~\cite{rfc4034}. In a DNSSEC-enabled zone, each NSEC record establishes a link between every two alphabetically consecutive domains within the zone. Upon receiving a query that does not exist within the zone, the ANS returns the NSEC record that contains the names that are alphabetically before and after the non-existent queried record name. In a \textit{zone enumeration} attack, an attacker can iteratively query all existing NSEC records and extract information about the domain names in the zone.

In order to mitigate \textit{zone enumeration} attacks, NSEC3~\cite{rfc5155} was introduced. In NSEC3, instead of returning the plaintext of the next and previous alphabetically closest domain names to the queried domain name, the ANS returns the hashes, which are alphabetically closest to the hash of the non-existent queried record~\cite{rfc5155}. However, in NSEC3 an attacker still can gather all the NSEC3 records within a zone and perform an offline dictionary attack~\cite{DNSCurveweb, goldberg2014nsec5}. The offline attack works by calculating the hash values of candidate names from a dictionary and comparing them against the hashes included in the collected NSEC3 records. Additional solutions, such as NSEC records with \textit{white lies}~\cite{rfc4470} and NSEC5~\cite{goldberg2014nsec5}, have been proposed to use real-time solutions to mitigate the longstanding problem of zone enumeration in NSEC and offline dictionary attack in NSEC3. However, they either require ZSK to be available on the ANSes of a zone~\cite{rfc4470} or introduce new keys on the ANSes~\cite{goldberg2014nsec5}, and their adoption remains limited. While the contents of DNS zones are not inherently confidential, extracting the entire domain names within a zone can reveal valuable information about the targeted domain (\eg existing servers or applications).

\textbf{Stale Records}: Another limitation in DNSSEC is the presence of stale, signed resource records. DNSSEC RRSIGs have an expiry window, determined by their \texttt{Inception} and \texttt{Expiration} fields. Stale records in DNSSEC come into existence when a signed resource record exists, and before its expiration (the time in the \texttt{Expiration} field has not yet been reached), a new resource record with an identical \texttt{name} but different \texttt{data} field gets signed. Although the resource record has been updated and a new, valid resource record is now available, the stale resource record is signed and has a valid, unexpired signature until the \texttt{Expiration} is reached. Stale resource records in DNSSEC are susceptible to replay, thereby enabling stale/false response injection. Replaying resource records can also be exploited to misdirect clients to non-optimal CDN nodes~\cite{hao2018end}. The absence of real-time and fresh signatures in DNSSEC enables such attacks.

\textbf{Expired Zone}: DNSSEC RRSIGs have a fixed expiry window, and DNSSEC-protected zones need to renew these signatures before the \texttt{Expiration} time. Failing to renew DNSSEC signatures may result in zone records being considered invalid, rendering the responses unacceptable to DNSSEC-validating resolvers. Zone records would thus become unreachable to clients that use validating resolvers.

\subsection{DNSCurve}
DNSCurve~\cite{curvedraft, DNSCurveweb} was proposed in 2009 as a backward-compatible solution to address the security, privacy, and amplification problems of DNSSEC. It uses authenticated encryption, where the public keys of ANSes are encoded and concatenated (as a subdomain) to the domain names of ANSes (\eg \texttt{``uz5jm...235c1.dnscurve.org''}). These concatenated public keys are 54 bytes long, including a hard-coded string `uz5’ added at the beginning of public keys, indicating support of DNSCurve by an ANS.

Similar to DNSSEC, in DNSCurve, Stage~1 is required to be secured separately. Also, in DNSCurve, resolvers do not signal clients regarding the use of DNSCurve in Stage~2~\cite{dnssecvs}. Hence, even if Stage~1 is secured, clients have no way of knowing that name resolution occurred securely using DNSCurve in Stage~2. In order to employ real-time authenticated encryption, DNSCurve requires private keys to be present on the nameservers of a zone. Therefore, when anycast is implemented by a zone owner for load balancing and enhancing performance, the private key needs to be present on all nameserver instances to facilitate online cryptographic operations~\cite{SchmidThirty}. The anycast instances are distributed across distinct geographical locations and administered in different regions, which the zone owner may not completely trust (\eg the root zone~\cite{IanaDNS}). Consequently, vulnerabilities of anycast server instances will impact the duplicated private key on said servers.

Aside from the replication of long-term secrets on nameserver instances and the absence of appropriate key distribution mechanisms to distribute keys among the nameserver instances in DNSCurve, the public keys of DNSCurve are appended as a subdomain to the nameserver names. Consequently, recursive resolvers typically obtain the nameserver keys from the nameserver of the parent zone. The nameserver records will be obtained securely only if the parent zones up to the root zone have also implemented DNSCurve and DNSCurve public keys have been retrieved securely. Therefore, we need to incorporate a DNSSEC-like trust anchor for DNSCurve to ensure secure communication with the root nameserver, securely obtain NS records and public keys of subordinate nameservers in the DNS hierarchy, and thereby securely transmit DNS messages. The absence of a properly defined trust model and chain of trust, by which the resolvers can trust the keys of nameservers in DNSCurve, is another problem of this scheme~\cite{SchmidThirty}.

In order to have a DNSSEC-like trust anchor in DNSCurve, the root nameservers are required to include their public keys in the recursive resolvers and duplicate their private keys on all of the nameserver instances for live cryptographic operations. ICANN, which is the entity responsible for managing the root nameserver, opted against replicating DNSCurve private keys across all nameserver instances to prevent the potential risks associated with exposing private keys on the nameserver instances~\cite{IanaDNS}. In contrast, DNSSEC can employ pre-signed resource records without requiring duplication of private keys on the nameserver instances within a zone.

\section{Threat Model and Required Properties}
\label{sec:threatandprops}
In this section, we detail the threat model of DNS resolution in Stage~2 and, based on the existing threats, define the required properties of ss2DNS to mitigate these threats. We also define the desired deployability-related properties for a DNS scheme in Stage~2 to facilitate widespread adoption.

\subsection{Threat Model}
\label{ssec:threats}
An adversary can mount off-path or on-path active and passive attacks in Stage~2. Active attacks enable security and availability threats, whereas passive attacks are often sufficient to compromise privacy.

\textbf{Security Threats:} Stage~2 threats can be posed by on-path or off-path adversaries. For example, an on-path adversary can inject false responses to a recursive resolver. An off-path adversary may inject false responses using techniques like the Kaminsky attack~\cite{KaminskyAttack}, inferring randomized values via side-channels~\cite{poisoningReloaded}, or exploiting IP fragmentation to avoid guessing attacks~\cite{herzberg2013fragmentation}. If these false responses are cached on a recursive resolver, these attacks also result in \emph{DNS cache poisoning}. Another type of security threat in Stage~2 is \emph{ANS replay} attack, in which an adversary captures previous responses from an ANS and replays them later. Even with DNSSEC, replay attacks can result in stale (false) response injection.

\textbf{Availability Threats:} The second category of active attacks in Stage~2 are ones that degrade or disrupt the name resolution availability by overloading ANSes. Such attacks can be mitigated in practice using CDNs, DoS detection, and/or rate-limiting techniques. There are also (D)DoS attacks that leverage the DNS infrastructure (\ie ANSes and recursive resolvers) to reflect and amplify traffic. Such attacks are prevalent among UDP-based DNS security schemes, including DNSSEC~\cite{DNSSECAmp, anagnostopoulos2013dns}.

\textbf{Privacy Threats:} Passive attacks in Stage~2 compromise the privacy of DNS queries and responses. Adversaries can collect information about the queries transmitted by a recursive resolver through eavesdropping, wherein metadata associated with queries (\eg source IP address, timestamp) belongs to the recursive resolver. Although the query metadata in Stage~2 does not belong to clients directly, the DNS query payload may contain client-related identifier fields such as EDNS Client Subnet (ECS)~\cite{ecs16}, or a query that can be linked to a specific client (\eg \textit{admin.example.com}), which results in gathering client-related information in Stage~2.

\subsection{ss2DNS Properties}
\label{ssec:properties}
We define security and privacy properties required to be satisfied in ss2DNS to mitigate the threats in Stage~2. Additionally, we define deployability and performance-related properties to avoid deployability barriers and facilitate widespread adoption of ss2DNS.

\textbf{Message Authentication}: False response injection can be performed by on-path and off-path attackers in Stage~2 (Sec.~\ref{ssec:threats}). ss2DNS provides message authentication and integrity to prevent unauthorized manipulation and injection of false responses.

\textbf{Anti-replay Protection}: In DNS schemes where responses are not freshly generated (\eg DNSSEC~\cite{ReplayStale, yan2008limiting}), responses can be replayed, resulting in vulnerability to stale-response injection. To mitigate response replay from previous interactions, we use a Time-Variant Parameter (TVP)~\cite{van2020computer} in ss2DNS to ensure message freshness.

\textbf{Avoid Duplicating Long-term Secret}: In secure DNS schemes that use real-time cryptographic operations (\eg encryption or signing), the private key is required to be present on the ANSes.  However, duplicating private keys across a zone’s nameservers increases the risk of key exposure due to targeted attacks on ANS instances. For example, DNSCurve requires private keys to be present on all ANS instances to securely transmit DNS messages~\cite{SchmidThirty}, which led to the root zone’s reluctance to adopt the scheme~\cite{IanaDNS}. To mitigate the exposure of long-term secrets within each zone and encourage widespread adoption, ss2DNS aims to satisfy this property.

\textbf{Confidentiality and Forward Secrecy}: Since DNS messages in Stage~2 are susceptible to eavesdropping, which may allow adversaries to gather client-related information (\eg ECS~\cite{ecs16}) or collect information about clients using a specific recursive resolver, ss2DNS provides response confidentiality to mitigate these privacy threats. Additionally, ss2DNS offers an optional query-encryption mode to further enhance privacy, as discussed in Section~\ref{ssec:resolution}. Moreover, in addition to response confidentiality, ss2DNS provides forward secrecy for responses to mitigate the privacy risk of future long-term key compromises.

\textbf{Mitigate Amplification}: Although UDP-based secure DNS schemes enhance efficiency, they are susceptible to reflection and amplification attacks. In ss2DNS, we aim to retain UDP-based resolution while limiting the amplification factor to a small constant.

\textbf{Failing closed}: In a secure DNS scheme, if message authentication fails or the cryptographic primitives used are not supported~\cite{DNSSECAgility}, DNS messages cannot be trusted and should be considered invalid. Adhering to the Safe-Defaults~\cite{van2020computer} principle, if at any point in the name resolution process of ss2DNS any verification fails, name resolution should be terminated and results discarded. Otherwise the provided properties cannot be trusted, potentially creating a false perception that the protocol is functioning flawlessly. Thus, by failing closed, ss2DNS effectively mitigates potential downgrading attacks that can bypass security validations within the protocol.

\textbf{Single Round-trip}: Since communications over the Internet are often preceded by a DNS query, any viable DNS scheme must minimize latency. Therefore, one of the key deployability and usability goals of ss2DNS is to achieve single round-trip query resolution over UDP, thereby reducing both network and overall delay in Stage~2. Issues related to UDP unreliability, such as erroneous or oversized segments, must thus be handled, as is true for Vanilla DNS as well.

\textbf{Established trust model}: Stage~2 DNS security schemes need to provide a \textit{trust model}, which refers to mechanisms that allow recursive resolvers to trust the keys used by nameservers. The web trust model is prevalent over the Internet, with billions of issued certificates~\cite{CFCT}. The web Public Key Infrastructure (PKI) has been used by Stage~1 schemes, such as DoH and DoT~\cite{jahromi2021comparative}. However, in Stage~2, the web PKI has been rarely used. We believe the reason for this is that TLS-based schemes (\eg DoT, DoH) are relatively expensive for Stage~2, and the root zone as a core authority within the Internet infrastructure is reluctant to rely on external entities (\eg CAs) in the web PKI as its trust anchors. For ss2DNS, we use a reverse-tree hierarchical (DNSSEC-like) trust model, which has been accepted and adopted by the root and TLDs within the DNS hierarchy.

\section{ss2DNS Technical Details}
\label{sec:specification}
In ss2DNS, cryptographic keys used for response authentication are signed in real-time on nameserver instances, without duplicating long-term private keys of zones. This design choice in ss2DNS prevents the duplication (copying) of \emph{precious} zone signing keys across hundreds of potentially untrusted \emph{nameserver instances} (the physically distributed server replicas deployed globally). The main premise here is to allow each (untrusted) nameserver instance to sign cryptographic parameters using its own unique short-term key and have that key being \emph{authorized} by a central key server constituting the main ANS of the zone. The central key server authorizes a nameserver's key by signing it, and revokes the key by refraining from renewing the signature. Such key signing can be implemented in automated fashion, allowing for very short key lifetimes (\eg few hours). This design fundamentally shifts the perception of replicated DNS zone server instances from the standard \emph{``logically centralized but physically distributed''} notion to a \emph{``delegated servers''} notion.\footnote{Not to be confused with DNS \emph{zone delegation}, where an entire DNS zone is delegated to other ANSes. The new delegation we are referring to in ss2DNS happens \emph{within} a zone.}

In what follows, we detail how ss2DNS operates, how a recursive resolver follows the chain of trust to verify the authorization of a server instance, and how query-response privacy can be added to ss2DNS without introducing extra network round-trips between the resolver and any nameservers.

\begin{figure}[ht!]
\centering
\includegraphics[width=0.85\linewidth]{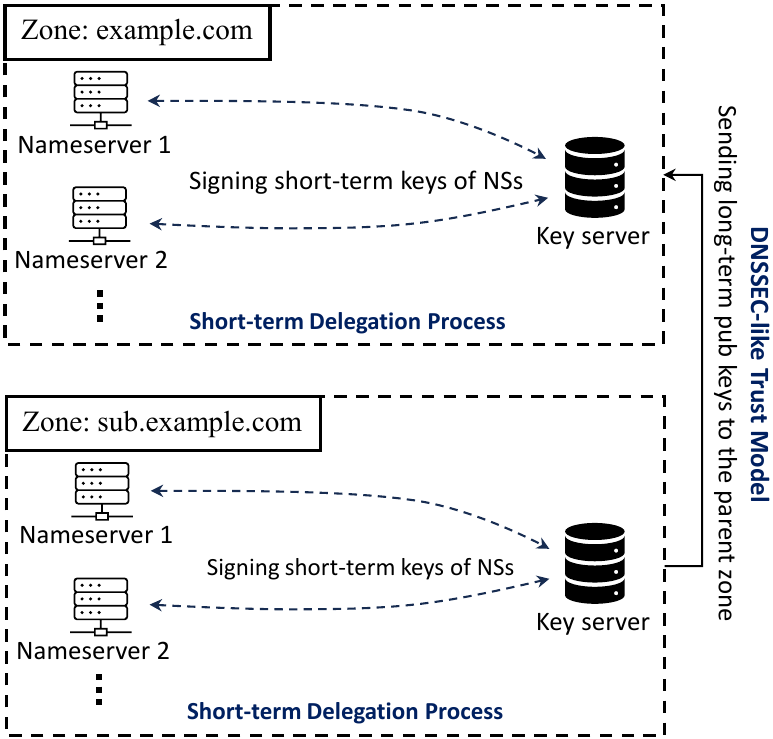}
\caption{Short-term delegation in ss2DNS and the trust model where each zone's public key is authenticated by its parent.}
\label{fig:zoneOverview}
\end{figure}

\textbf{Nameserver delegation.} Figure~\ref{fig:zoneOverview} shows a zone in ss2DNS. Each zone has a \textsl{central key server} (or \textsl{``key server''} for short), which is trusted by, and under direct control of, the zone administrator. Its purpose is to store the long-term signing key of the zone and delegate nameserver instances within the zone by signing their keys. This delegation authorizes nameserver instances to respond to queries.

\textbf{Trust model.} A reverse-tree chain of trust (analogous to DNSSEC) is used in ss2DNS, where the public component of the long-term signing key of each zone (\ie the verification key) is placed in the parent zone. The public component of the long-term key of the root zone is installed in DNS resolver software as a trust anchor.

\textbf{Real-time integrity protection of DNS responses.} 
With access to a zone's long-term verification key (from the parent zone or hard-coded as a trust anchor), a recursive resolver querying a nameserver instance first verifies the authenticity of the nameserver's short-term key. This short-term key is then used to establish a symmetric key between the nameserver and the resolver, enabling the transfer of an authenticated, encrypted DNS response.

\textbf{Additional feature: query confidentiality.} In addition to response confidentiality, ss2DNS offers query confidentiality (for privacy) as an optional feature. It operates in two modes: \emph{no-privacy} and \emph{privacy-enforcing}. The former provides confidentiality for DNS responses only, while the latter protects confidentiality of both queries and responses. A notable challenge in the privacy-enforcing mode is that the recursive resolver must first obtain the nameserver's short-term key from the nameserver itself (which requires one round-trip) and then use it to encrypt the query and send it to the nameserver (requiring a second round-trip). This doubling of round-trips would significantly hinder the practical adoption of ss2DNS.

To avoid requiring an additional round-trip, we use two different symmetric keys: one to encrypt the query and the other to encrypt the response. To establish the query key with a nameserver instance, the resolver obtains all needed information from the parent zone's nameserver. When it transmits the encrypted query, it sends with it its own Diffie-Hellman (DH) agreement key (the $g^a$) in the same transmission. Upon receiving this, the nameserver instance decrypts the query, generates the response key, encrypts the response with it, and sends it along with its freshly-generated DH agreement key (the $g^b$), which it signs with its own short-term key (itself was signed by the long-term zone key).\footnote{We use modular exponentiation (modp) for exposition, but expect EC to be used in practice.} Note that while the query-encryption key is now accessible to all nameserver instances (unlike the response-encryption key), an adversary compromising that key does not impact the confidentiality and authenticity of the responses (as they use a different key).

\textbf{Forward Secrecy.} ss2DNS implements a half-static DH key agreement approach for queries to enable DNS resolution within a single round-trip. Thus, forward secrecy is not provided for queries. However, ss2DNS provides forward secrecy for responses, as the nameserver uses the resolver’s ephemeral key (included in the query) and generates a new ephemeral key for each response transmission.

\subsection{Zones in ss2DNS}
\label{ssec:zones}
In each zone, there is a key server trusted by the zone owner (Fig.~\ref{fig:zoneOverview}), and there are other nameserver instances that may not be completely trusted by the zone owner. The nameserver instances can be nameservers under the control of the zone administrator, or globally distributed nameserver instances managed by a CDN service provider, which are not directly controlled by the nameserver administrator, and do not have access to the long-term private key of the zone.

\lstset{
language=C,
basicstyle=\ttfamily\small\color{textcolor},
columns=fullflexible,
frame= lines,
keywordstyle=\color{magenta},
commentstyle=\color{myTeal},
backgroundcolor=\color{white},
caption=Short-term nameserver signing key structure,
keywords=[2]{Pubkey},
keywords=[3]{SignatureScheme},
keywords=[4]{Signature},
keywords=[5]{SignatureScheme},
keywords=[6]{uint32},
keywordstyle={\color{blue!80!black}},
keywordstyle=[2]{\color{blue!80!black}},
keywordstyle=[3]{\color{blue!80!black}},
keywordstyle=[4]{\color{blue!80!black}},
keywordstyle=[5]{\color{blue!80!black}},
keywordstyle=[6]{\color{blue!80!black}},
numberstyle=\footnotesize\color{codegray},
numbers=left,                    
numbersep=-8pt,
}

\begin{figure}
    \centering
    \begin{lstlisting}[label=STKstruct,captionpos=b]
    struct {
        struct {
            int inception;
            int expiration;
            Pubkey STK_public_key;
            int nameserver_ID;
            int zone_level;
        } short_term_key_structure;
        Signature signature;
    } Signed_short-term_key_structure;
    \end{lstlisting}
    \vspace{-20pt}
\end{figure}

Table~\ref{tab:symbols} lists the symbols used for specifying keys, zones, and nameservers. A zone with level $l$ in the DNS hierarchy has a long-term signing key ($\underline{\text{w}}_l$), stored on the key server of the zone. By a secure but unspecified means, the nameservers and the key server within a zone must be able to mutually authenticate each other and confidentially exchange messages.\footnote{Different CDN providers may use different methods to ensure secure interactions with their edge servers, and we do not impose any specific constraints on these options.} A nameserver with ID $i$ in a zone with level $l$ in the DNS hierarchy generates a fresh short-term signing key structure ($\omega_l^i$). Subsequently, the nameserver sends its short-term public key structure ($\omega_l^i$) through the described secure channel to the key server of the zone. The key server verifies the short-term key structure and its origin nameserver and, upon successful validation, signs the short-term public key structure of nameservers ($\omega_l^i$) using the zone's long-term signing key ($\underline{\text{w}}_l$). Finally, the key server returns the signed short-term key structure to the nameserver.

As Listing~\ref{STKstruct} shows, the short-term signing key of a nameserver consists of a key value plus 4 attributes. \texttt{inception} and \texttt{expiration} values indicate the lifetime of the short-term key structure. \texttt{STK\_public\_key} is the short-term signing public key of a nameserver ($\omega^i_l$), and \texttt{nameserver\_ID} indicates the unique ID of a nameserver within a zone. Finally, \texttt{zone\_level} field indicates the level of the zone within which this short-term key is signed. These five fields constitute the \texttt{short\_term\_key\_structure}, which will be signed by the long-term key of a zone. The signed structure with the included \texttt{Signature} field then forms the \texttt{Signed\_short-term\_key\_structure}.

\begin{table}[hb!]
    \centering
    \caption{Symbols used in the abstract description of ss2DNS operation: The top four are asymmetric keys, and the bottom three are ownership annotation. The asymmetric key symbols (top four) will represent the public component of the key (agreement or signature verification), and for their private component (agreement or signing), the symbol is underlined.}
    \begin{tabular}{@{}cl@{}} 
        \thickhline
        \multicolumn{1}{@{}c@{}}{Symbol} & \multicolumn{1}{c}{Meaning} \\ \hline
        $\mathcal{A}$, $\underline{\mathcal{A}}$ & Long-term public, private agreement key \\ 
        $\Lambda$, $\underline{\Lambda}$ & Short-term public, private agreement key \\ 
        $\text{w}$, $\underline{\text{w}}$ & Long-term verifying, signing key \\  
        $\omega$, $\underline{\omega}$ & Short-term verifying, signing key \\ \hline
        $r$ & Unique random number \\ \hline
        $l$ & Zone level in the DNS hierarchy (subscript) ($0 \leq l$) \\ 
        $i$ & Nameserver ID number (superscript) ($0 \leq i$)\\ 
        $R$ & Recursive resolver (superscript)\\ \thickhline
    \end{tabular}
    \label{tab:symbols}
\end{table}

\begin{figure}[ht!]
\centering
\includegraphics[width=0.9\linewidth]{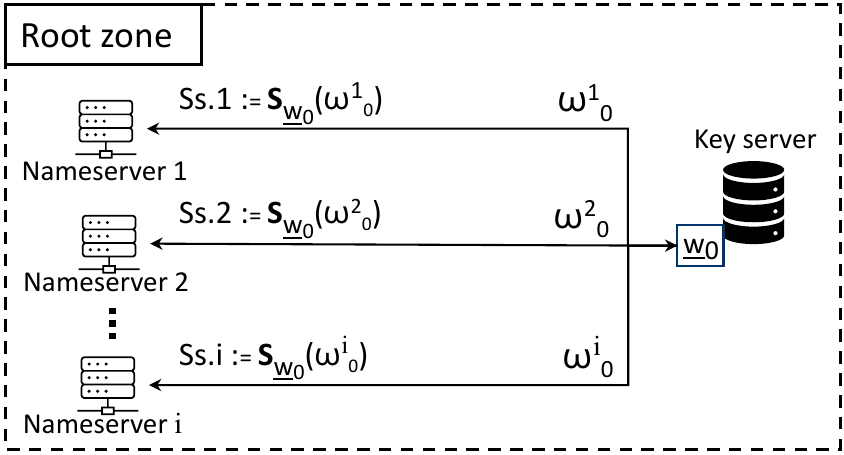}
\caption{The process of signing the nameserver short-term key structures by the long-term signing key of the zone.}
\label{fig:signingProcess}
\end{figure}

For instance, Fig.~\ref{fig:signingProcess} illustrates the process of signing short-term key structures in the root zone. As the top arrow shows, \textit{Nameserver 1} generates a short-term key structure ($\omega_0^1$), and sends it to the root zone's key server. Upon securely receiving the short-term public key structure of \textit{Nameserver~1} ($\omega_0^1$), and validating the key structure and authenticating the nameserver, the zone's key server signs the short-term public key structure of the nameserver ($Ss.1\:=\:S_{\underline{w}_0}(\omega_0^1)$) using the long-term signing key of the zone ($\underline{\text{w}}_0$). Subsequently, the key server securely sends the signed short-term key of the nameserver to \textit{Nameserver 1}. The signed structure of short-term keys of nameservers have a validity period that specifies their lifetime. The signed short-term key structures have a brief lifetime (\eg hours to days). Thus, short-term key structures minimize the threat and exposure of compromised keys and ensure implicit revocation of nameserver keys in short time intervals.

\begin{table*}
    \centering
    \caption{List of functions used in ss2DNS}
    \begin{tabular}{@{}lll@{}} 
        \thickhline
         \multicolumn{1}{c}{Function} & \multicolumn{1}{c}{Used to} & \multicolumn{1}{c}{Symbol} \\ \hline 
         Symmetric authenticated encryption & Encrypt message $m$ with key $a$ & $E_{a}(m)$ \\ 
         Symmetric authenticated decryption & Decrypt message $m$ with key $a$& $D_{a}(m)$ \\ 
         Signature generation & Sign message $m$ with key $a$& $S_{a}(m)$ \\ 
         Signature verification & Verify signature on message $m$ with key $b$ & $V_b(m , S_a(m))$\\
         Key establishment & Produce DH key using private key $\underline{A}$ and public $B$ & $DH(\underline{A}, B)$\\
         Generate ephemeral key pair& Generate ephemeral agreement keys & $(\underline{A}, A) := GenDH()$\\ \thickhline
    \end{tabular}
    \label{tab:functions}
\end{table*}

\begin{table*}
    \centering
    \caption{List of keys used in ss2DNS (A-encryption and A-decryption are authenticated functions)}
    \begin{tabular}{cll}
        \thickhline
        \multicolumn{1}{c}{Label} & \multicolumn{1}{c}{Key type} & \multicolumn{1}{c}{Used to} \\ \hline
        \multicolumn{3}{c}{Zone Keys} \\ \hline
        $\underline{\mathcal{A}}_l$ & Long-term zone \textbf{private agreement key} & Establish shared secret for query A-decryption \\ 
        $\mathcal{A}_l$ & Long-term zone \textbf{public agreement key} & Establish shared secret for query A-encryption \\ 
         $\underline{\text{w}}_l$ & Long-term zone \textbf{signing key} (private) & Sign short-term nameserver keys \\ 
         $\text{w}_l$ & Long-term zone \textbf{verifying key} (public) & Verify short-term nameserver keys \\ \hline
         \multicolumn{3}{c}{Nameserver Keys} \\ \hline
         $\underline{\omega}^i_l$ & Short-term nameserver \textbf{signing key} (private) & Sign ephemeral session agreement keys \\ 
         $\omega^i_l$ & Short-term nameserver \textbf{verifying key} (public) & Verify ephemeral session agreement keys \\ 
         $\underline{\Lambda}^i_l$ & Ephemeral nameserver \textbf{private agreement key} & Establish shared secret for response A-encryption \\ 
         $\Lambda^i_l$ & Ephemeral nameserver \textbf{ public agreement key} & 
         Establish shared secret for response A-decryption \\ \hline
         \multicolumn{3}{c}{Resolver Keys} \\ \hline
         $\underline{\Lambda}^R$ & Ephemeral resolver \textbf{private agreement key} & Establish shared secret for query and response \\ 
         $\Lambda^R$ & Ephemeral resolver \textbf{public agreement key} & Establish shared secret for query and response \\ \thickhline
    \end{tabular}
    \label{tab:keys}
\end{table*}

Before the expiration of the current signed key structure, the nameserver instances generate a new short-term signing key structure. Subsequently, this newly generated key structure is transmitted to the zone's key server via an authenticated and encrypted channel to be signed. If the nameservers within a zone do not renew their short-term signing key structures prior to the expiration of the current key, the resolvers cannot validate the responses after expiration of the current key, and the ss2DNS resolution fails. The long-term keys in ss2DNS are stored securely on the key server of each zone. Thus, the attack surface of the long-term keys is significantly smaller compared to the short-term key structures, which are stored on the nameserver instances.

Aside from the long-term signing key of each zone ($\text{w}_l$), which is stored on a key server within each zone, there is another long-term agreement key associated with each zone ($\mathcal{A}_l$). See Table~\ref{tab:keys} for a complete list of ss2DNS keys. To provide confidentiality of DNS queries, resolvers need to have access to a public agreement key from the nameservers. The retrieval of this key from the nameserver requires an additional round-trip, violating our desired single round-trip policy (Sec.~\ref{ssec:properties}). In order to satisfy the single round-trip and confidentiality properties at the same time, each zone with level $l$ contains another long-term initial agreement key ($\mathcal{A}_l$). Unlike the zone's private long-term signing key ($\underline{\text{w}}_l$), which is stored only on the key server within each zone, the private long-term initial agreement key ($\underline{\mathcal{A}}_l$) is transferred to all the nameserver instances within each zone. In ss2DNS, when a zone generates $\mathcal{A}_l$, it is required to transmit it to the parent zone along with the zone's long-term signing key ($\text{w}_l$). Then, $\mathcal{A}_l$ is used to provide confidentiality of DNS queries, as we explain next in Sec.~\ref{ssec:resolution}. Based on the decision of resolvers on the privacy level of queries, they can use the long-term agreement key of zones for query encryption.

\begin{figure}[ht!]
\centering
\includegraphics[width=1\linewidth]{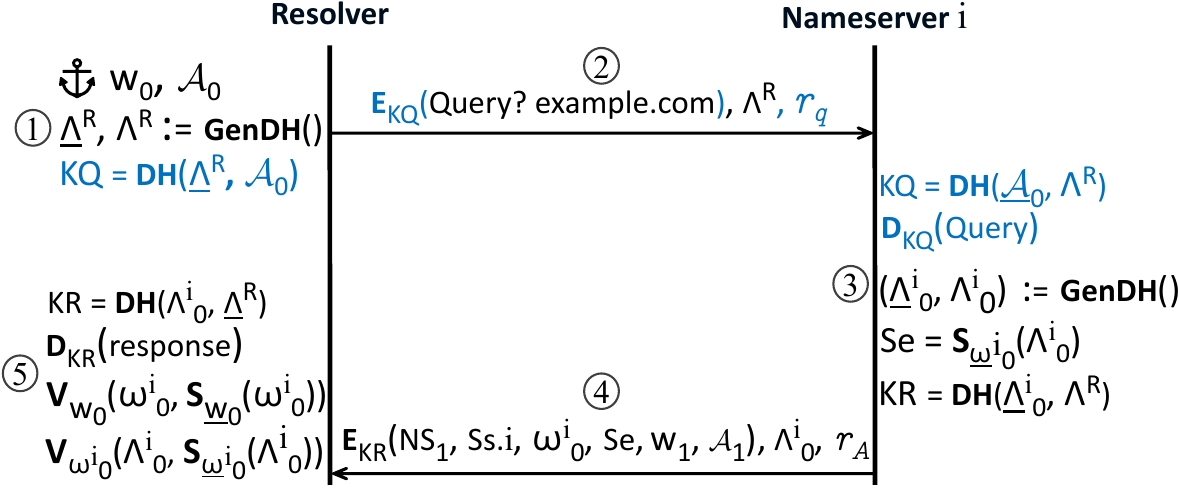}
\caption{ss2DNS query resolution from a resolver to nameserver $i$ of the root zone. The steps in black occur in both no-privacy and privacy-enforcing modes; steps in blue only occur in privacy-enforcing mode (query encryption).}
\label{fig:TDiagram}
\end{figure}

\subsection{Name Resolution in ss2DNS}
\label{ssec:resolution}
In ss2DNS, ANS $i$ within a zone with level $l$ has two keys (see Sec.~\ref{ssec:zones}): one short-term for signing ($\omega^i_l$) and one long-term for key agreement ($\mathcal{A}_l$). The key $\omega^i_l$ is signed by the long-term signing key of the zone ($\underline{\text{w}}_l$), which is stored on the zone's key server. A ss2DNS resolver has access to the long-term public keys of the root ($\text{w}_0$, $\mathcal{A}_0$) as trust anchors.

In ss2DNS, resolvers can operate in two modes: privacy-enforcing and no-privacy. Based on the privacy sensitivity of queries (\eg when ECS~\cite{ecs16} is included) or per client (stub resolver) request, they have the option to encrypt the transmitted queries in the privacy-enforcing mode. We use the notation in Table~\ref{tab:functions} to represent cryptographic functions. Also, Table~\ref{tab:keys} classifies the keys within ss2DNS based on their owner entities. The private part of an asymmetric key pair is expressed using underlined letters (\eg $\underline{\mathcal{A}}$ is a private key and $\mathcal{A}$ is its corresponding public key).

\textbf{No-privacy mode.} Figure~\ref{fig:TDiagram} illustrates the process of name resolution in ss2DNS when resolving a domain name from nameserver $i$ of the root zone. The steps written in black occur when a resolver is in the no-privacy mode. In Step~1, the resolver generates an ephemeral agreement key pair ($\underline{\Lambda}^R,\: \Lambda^R$). Subsequently, in Step~2, the resolver transmits the plaintext query (\textit{Query? example.com}) with the resolver's ephemeral public key ($\Lambda^R$) to nameserver $i$. Upon receiving the query and looking up the response in Step~3, nameserver $i$ generates an ephemeral agreement key pair ($\underline{\Lambda}^i_0,\: \Lambda^i_0$). Then, the ephemeral public key of the nameserver ($\Lambda^i_0$) is signed ($Se\:=\:S_{\underline{\omega}^i_0}(\Lambda^i_0)$) using the short-term signing key of the nameserver ($\underline{\omega}^i_0$). At this point, the nameserver generates a master key ($KR$) using DH key agreement with the ephemeral private key of the nameserver ($\underline{\Lambda}^i_l$) and the ephemeral public key of the resolver ($\Lambda^R$). The generated master key and the fresh random number ($r_A$) are used as inputs of a Key Derivation Function (KDF) to derive the encryption key of the response. In addition to the standard DNS response, additional cryptographic parameters are appended to the response prior to encryption.

As Fig.~\ref{fig:TDiagram} shows, in this example, the resolver queries the root zone nameserver for `\textit{example.com},' and the root zone is not authoritative for providing the final response to this query. Therefore, nameserver $i$ within the root zone returns a nameserver `NS' record for the TLDs at level~$1$ in the DNS hierarchy. As demonstrated in Step~4, the nameserver uses the master key ($KR$) derived in Step~3 with a fresh random number ($r_A$) and a KDF to encrypt the `NS\_1' record of the TLD. Additionally, the nameserver appends the short-term key structure ($\omega^i_0$) of the nameserver with its corresponding signature ($Ss.i$). The signature is generated by the long-term key of the zone on the key server within the zone ($Ss.i\:=\:S_{\text{w}_l}(\omega^i_l)$), as described in Sec.~\ref{ssec:zones}. Moreover, the signature of the ephemeral key of the nameserver ($Se\:=\:S_{\underline{\omega}^i_l}(\Lambda^i_l)$) is appended to the message before encryption. In this example, the $NS_1$ in the response belongs to a delegated zone, so the long-term signing ($\text{w}_1$) and query agreement key ($\mathcal{A}_1$) of the TLD are also appended to the response message. These long-term keys of the TLD will be used when the resolver sends queries to the TLD nameservers. After encrypting the DNS response with additional cryptographic signatures and keys, the nameserver appends the public ephemeral key ($\Lambda^i_l$) along with the random number ($r_A$) used for encrypting the response. Subsequently, the nameserver sends the response to the resolver in Step~4.

Upon receiving the response, in Step~5, the resolver generates the master key ($KR$) using the public ephemeral key of the nameserver ($\Lambda^i_0$) and the resolver's ephemeral private key ($\underline{\Lambda}^R$). It then decrypts the message within the response, and verifies the signature of the short-term public key structure of the nameserver ($S_{\underline{\text{w}}_0}(\omega^i_l)$) using the long-term signing key of the root zone ($\text{w}_0$). Next, the resolver verifies the signature of the ephemeral nameserver public key ($S_{\underline{\omega}^i_0}(\Lambda^i_0)$), which was used to encrypt the response. If the decryption or signature verifications fail, the response is considered invalid and discarded. Otherwise, if all checks in Step~5 complete successfully, the resolver caches and uses the DNS response.

After securely resolving the NS record of TLD from the root zone, the resolver has access to the long-term public keys of the `\textit{.com}' zone (\ie $\text{w}_1,\:\mathcal{A}_1$). The resolver is now able to repeat the same steps for resolving Second-Level Domain (SLD) NS records. When the resolver reaches the nameserver authoritative for the queried record, it repeats the same steps. However, the response does not contain the long-term keys of the child zone (\ie $\text{w}_{l+1},\:\mathcal{A}_{l+1}$), as at that point, the resolver does not need to traverse other subordinate zones.

\textbf{Privacy-enforcing mode.} To resolve DNS records in one round trip while encrypting queries, we separated the long-term zone key used for providing security and privacy properties of queries from the long-term key used for responses. The blue steps in Fig.~\ref{fig:TDiagram} are used in the privacy-enforcing mode. After generating the ephemeral key pair, the resolver generates a master key using DH key agreement ($GenDH()$) with the root zone's initial agreement public key ($\mathcal{A}_0$) and the resolver's ephemeral agreement private key ($\underline{\Lambda}^R$). The generated master key ($KQ$) is then used for query confidentiality (and integrity). In Step~2, the resolver uses authenticated encryption with a key derived from ($KQ$) to protect the integrity and confidentiality of the query. The authenticated encryption herein uses random numbers used once as TVP to ensure freshness of encryption keys~\cite{NISTKDF}. The resolver then transmits the encrypted query, along with the resolver's ephemeral public key ($\Lambda^R$) and the random number ($r_q$) used in derivation of the encryption key, to nameserver $i$ of the root zone.

The nameserver $i$ receives the encrypted query with the resolver's ephemeral public key and the random number from Step~2. The nameserver generates the same master key ($KQ$), using DH key agreement with the resolver's ephemeral public agreement key ($\Lambda^R$) and the root zone's private long-term agreement key ($\underline{\mathcal{A}}_0$). The nameserver uses the generated master key and the received nonce from the resolver to decrypt the query. The next steps after decrypting the query are the same as the steps explained in the no-privacy mode.

\subsection{Caching}
The caching mechanism for standard DNS records remains the same in ss2DNS. The resource records are transmitted as authenticated and encrypted messages. After decryption and verification, they will be treated as Vanilla DNS messages. Caching the long-term keys of the zones in ss2DNS is essential to achieve a comparable performance to Vanilla DNS. Otherwise, each time a new record needs to be resolved by a resolver, the resolver needs to traverse the DNS hierarchy to obtain the long-term keys of the intended zone to securely resolve the query. Regarding the period for which long-term keys are cached in ss2DNS by resolvers, caching for long- and short-term durations have similar advantages and drawbacks as DNSSEC keys (Sec.~\ref{sssec:DNSSECFlow}).

The long-term signing key in ss2DNS is stored on a trusted key server within each zone and not used directly in the interaction of nameservers and resolvers. With that in mind, caching long-term keys associated with zones for periods longer than DNS record TTL values is unlikely to raise security concerns while providing performance benefits. For example, the keys of a zone can be cached for 24 hours, and whenever the resolver intends to resolve a query from the nameservers within the caching period, the cached keys can be used without requiring communication with the parent zones to obtain the long-term keys of the zone. A practical approach for setting the caching time of the zone keys is to set the caching time of the long-term keys of the zones up in the DNS hierarchy (\eg root or TLDs) relatively longer compared to their subordinate zones. In this manner, when a resolver wants to resolve a domain name within a given zone, if the long-term keys of the intended zone are not cached, the resolver does not need to traverse the entire DNS hierarchy to obtain them, since there is a greater likelihood that the long-term keys for higher-level zones have been previously cached.

\subsection{Updating Records and Keys}
\label{sec:update}
\textbf{Standard DNS records}: As DNS messages in ss2DNS are securely sent using the original DNS zone files, the process of updating zone file records remains the same as Vanilla DNS.

\textbf{Updating Short-term keys ($\omega^i_l$)}: As such keys have short lifetime, nameservers need to obtain a new signed short-term key structure before the expiration of the current one.

\textbf{Updating long-term zone agreement keys ($\mathcal{A}_l$)}: Since long-term zone keys are used to establish the ss2DNS trust model, the process of updating long-term keys requires considerations to avoid name resolution failures. For updating the long-term agreement key ($\mathcal{A}_l$) of a zone with level $l$, the zone administrator initially generates a new agreement key $\mathcal{A}_l$ on its key server and sends it its zone nameservers, so that they can decrypt incoming queries encrypted using the new key. In the next step, the zone owner removes the old key from the parent zone and adds the new $\mathcal{A}_l$ to the parent zone using the OOB authenticated channel between the zones.\footnote{In DNSSEC domain registrars typically provide web interfaces for adding/removing keys in parent zones.} After waiting long enough for the old $\mathcal{A}_l$ to be removed from the caches of resolvers, the zone owner removes the old agreement key from its nameservers.

\textbf{Updating long-term zone signing keys ($\text{w}_l$)}: Updating long-term signing keys is similar to updating KSKs in DNSSEC, where three update methods exist~\cite{rfc7583}. However, for updating the zone signing keys ($\text{w}_l$) in ss2DNS, we use a customized approach, which is similar to the double-DS method in DNSSEC. This method is the most efficient regarding the number of additional bytes added to the responses during the long-term zone signing key updates.

For updating the long-term zone signing key, denoted as $\text{w}_l$ (old), to the new key, denoted as $\text{w}_l$ (new), the zone owner first adds $\text{w}_l$ (new) to the parent zone. At this point, the parent zone publishes both the old and new keys, and the zone owner waits for enough time to ensure the expiration of $\text{w}_l$ (old) from the resolver caches, and the $\text{w}_l$ (new) is cached alongside the old key in the caches of resolvers. Next, the zone owner removes the $\text{w}_l$ (old) from its zone and starts using $\text{w}_l$ (new) for signing the short-term key structures. Following this step, the zone owner waits for enough time, ensuring the expiration of short-term key structures signed by $\text{w}_l$ (old) in its zone. Finally, the zone owner removes the $\text{w}_l$ (old) from the parent zone, and the process is complete.

\section{ss2DNS Evaluation}
\label{sec:specification2}
This section provides a discussion into the amplification factor in ss2DNS, and compares ss2DNS with DNSSEC and DNSCurve. Next, we describe a PoC implementation of ss2DNS, followed by a comparative performance evaluation.

\subsection{Amplification Factor}
\label{ssec:amplification}
As explained in Sec.~\ref{ssec:properties}, it is crucial for ss2DNS to resolve queries in a single round-trip. There are trade-offs associated with a single round-trip, and amplification is one of the important aspects to consider. One of the schemes with a notable bad reputation regarding traffic amplification in Stage~2 is DNSSEC. Although the amplification factor in DNSSEC can theoretically exceed $100\times$, the empirically observed average amplification factor for queries of type \texttt{ANY} for TLDs in DNSSEC 2014 was approximately $47\times$~\cite{DNSSECAmp}. The queries of type \texttt{ANY} often result in a greater amplification factor. When an attacker abuses the \texttt{ANY} queries to target a DNSSEC-enabled nameserver, the nameserver returns any type of resource records available on the nameserver for the given domain name in response. In a DNSSEC-protected zone, in addition to the resource records, the nameserver also returns the RRSIGs associated with each resource record. Therefore, relative to the number of resource records included in the response, a DNSSEC-enabled server returns RRSIGs, which results in a greater amplification of traffic.

In ss2DNS, regardless of DNS record type or the number of records in the response, the number of bytes added to the response for encryption and authentication are constant (Figure~\ref{fig:qeuryresp}). The reason is that, unlike DNSSEC, for each DNS record, a separate signature is not required. Consequently, the amplification factor in ss2DNS is restricted and cannot be abused for considerable amplifications in DDoS attacks. With Elliptic Curve Digital Signatures (ECDSA) and NaCl cryptography~\cite{bernstein2009cryptography}, the number of additional bytes for a non-delegating response is $\sim$245 bytes, and for a delegating response $\sim$310 (see Sec.~\ref{sec:implementation}). The number of added bytes by ss2DNS can be further decreased (Sec.~\ref{sec:discussion}). Compared to DNSSEC, which can add thousands of bytes to a query response of type \texttt{ANY}, ss2DNS only adds a limited number of bytes for authentication and encryption.

\subsection{Comparative Analysis: DNSSEC, DNSCurve}
Compared to DNSSEC, which only provides message authentication to DNS responses, ss2DNS provides real-time authenticated encryption for encrypting DNS queries and responses, thereby providing both confidentiality and message authentication. Therefore, ss2DNS does not require NSEC-like records~\cite{rfc4034, rfc5155, goldberg2014nsec5} for negative responses, and regular non-existent domain (NXDOMAIN) responses can be transmitted securely. As explained in Sec.~\ref{ssec:amplification}, unlike DNSSEC, which is susceptible to significant traffic amplification rates, ss2DNS responses contain only a fixed number of additional bytes. Besides, unlike DNSSEC, which is susceptible to replay attacks, ss2DNS prevents replay between different sessions due to the use of ephemeral agreement keys. Additionally, the inclusion of a TVP ensures message freshness within a session. Consequently, even if multiple queries are sent using the same ephemeral key, the queries and responses cannot be replayed within the same session. Another difference between DNSSEC and ss2DNS is that DNSSEC requires separate queries to obtain the DNSKEY records from a zone's nameservers. Although both queries can be transmitted simultaneously and the delay would remain the same, in ss2DNS the keys are appended as part of the response, and one less query is required. Finally, DNSSEC requires modifications to the zone files, while in ss2DNS the zone files remain unchanged, which saves the administrative time of updating zone signatures.

Now, consider DNSCurve~\cite{DNSCurveweb, curvedraft} in the context of key management. DNSCurve lacks a defined mechanism for distributing nameserver keys across anycast instances. Furthermore, it requires long-term keys to be stored on nameserver instances, exposing them to attacks and preventing adoption by the root zone~\cite{IanaDNS}. To address these issues, ss2DNS introduces a delegation approach where a key server within the zone signs short-term key structures for nameserver instances. ss2DNS thus avoids duplicating long-term secrets and provides means for distributing the keys of nameserver instances within a zone.

DNSCurve~\cite{curvedraft} does not provide forward secrecy for queries or responses, as the ANS’s public key is not ephemeral. ss2DNS implements a half-static DH approach for queries to enable DNS resolution within a single round-trip, and therefore does not provide forward secrecy for queries either. However, ss2DNS provides forward secrecy for responses. Finally, DNSCurve does not provide a chain of trust in the DNS hierarchy. Thus, resolvers cannot validate the authenticity of an NS record that contains a public key, rendering DNSCurve susceptible to false nameserver injections~\cite{SchmidThirty, IanaDNS}. In ss2DNS, the long-term keys of zones establish a DNSSEC-like chain of trust up to the root, making it resilient to the aforementioned attack.

\subsection{Proof-of-Concept (PoC) Implementation}
\label{sec:implementation}
We built a PoC of ss2DNS for testing its performance in practice, and for comparison with other DNS schemes. The prototype consists of two parts: ANS-side and resolver-side.

\begin{figure*}[ht!]
\centering
        \subcaptionbox{Query}
        {\includegraphics[width=0.15\textwidth]{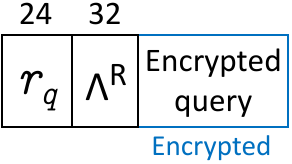}\label{fig:Query}}
\hspace{30pt}
        \subcaptionbox{Response}
        {\includegraphics[width=0.53\textwidth]{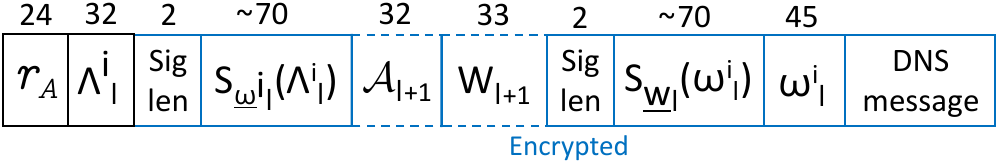}\label{fig:Response}}
    \caption{Query and response format in ss2DNS (the dashed boxes are only included in delegating responses).}\label{fig:qeuryresp}
\end{figure*}

\textbf{ANS-side.} So as to implement the encryption and decryption functions, we modified the DNS library used in \texttt{CoreDNS}~\cite{CoreDNS}. As demonstrated in Fig.~\ref{fig:qeuryresp} (b), the nameserver adds its short-term public key structure ($\omega^i_l$) with its signature generated by the zone's key server ($S_{\underline{\text{w}}_l}(\omega^i_l)$). Additionally, the signature of the ephemeral public agreement key ($S_{\underline{\omega}^i_l}(\Lambda^i_l)$) is added to the response prior to encryption. Finally, the public ephemeral key of the nameserver ($\Lambda^i_l$) and the random number ($r_A$) used to encrypt the response are added to the response. The dashed boxes represent the long-term keys associated with the child zone ($\mathcal{A}_{l+1},\:\text{w}_{l+1}$), and are added when the response is referring to a delegated zone. In responses to the queries for which a nameserver is authoritative, the dashed boxes are omitted. We used ECDSA with curve P-256 and SHA256~\cite{ECDSA} for signing and verifying the signatures, and NaCl~\cite{bernstein2009cryptography} library for authenticated encryption.

\textbf{Resolver-side.} The resolver encrypts DNS queries (Fig.~\ref{fig:qeuryresp} (a)) using NaCl-based~\cite{bernstein2009cryptography} authenticated encryption and sends the encrypted queries alongside the freshly generated ephemeral key of the resolver ($\Lambda^R$) and the random number ($r_q$), used for query encryption. Upon receiving an encrypted ss2DNS response, the resolver extracts the random number and public key from the message and decrypts the encrypted part. Subsequently, the resolver parses and extracts the included keys, signatures, and the DNS message from the response. The resolver initially verifies the digital signatures of the nameserver's short-term key structure and ephemeral key. If the signatures are verified successfully, the resolver proceeds to process the DNS response; otherwise, the response is discarded. To implement the resolver-side in ss2DNS, we modified \texttt{q}~\cite{exDNS}, which is a similar DNS resolution tool to \texttt{dig}, implemented in Go.

\newcommand{\abdouwidth}{1.3}

\begin{figure}[!ht]
    \centering
    \begin{subfigure}[t]{0.27\textwidth}
    %\centering
    \hspace*{-0.15cm}
    \begin{tikzpicture}[scale=1]
        \begin{axis}[width=\textwidth,
            y label style={at={(axis description cs:0.15,0.5)}},
            ymajorticks=true,
            xlabel={Time (ms)},
            x label style={at={(axis description cs:0.5,0)}},
            xmode = log,
            ylabel={CDF},
            xmin=0, xmax=400,
            ymin=0, ymax=1,
            ytick={0,0.2,0.4,0.6,0.8,1},
            legend style={
            at={(-0.12,1.1)},      
            anchor=south west,     
            legend columns=3,  
            font=\small
            },
            line width=0.4pt,
]

\addplot[line width=\abdouwidth\pgflinewidth, solid, color= green, mark size=1pt]
  coordinates {\input{Total-Latency/TotalUDPDump1.txt} };

\addplot[line width=\abdouwidth\pgflinewidth, densely dashdotted, color= violet, mark size=1pt]
  coordinates {\input{Total-Latency/TotalsecPlusNoPriv1.txt} };

\addplot[line width=\abdouwidth\pgflinewidth, dashdotted, color= orange, mark size=1pt]
  coordinates {\input{Total-Latency/TotalPlus1.txt} };

\addplot[line width=\abdouwidth\pgflinewidth, dotted, color= gray, mark size=1pt]
  coordinates {\input{Total-Latency/TotalPersistentTCP1.txt} };

\addplot[line width=\abdouwidth\pgflinewidth, dashed, color= blue, mark size=1pt]
  coordinates {\input{Total-Latency/TotalPreSECDump1.txt} };

\addplot[line width=\abdouwidth\pgflinewidth, dashdotdotted, color= cyan, mark size=1pt]
  coordinates {\input{Total-Latency/TotalLiveSecDump1.txt} };

\addplot[line width=\abdouwidth\pgflinewidth, dashdotdotted, color= purple, mark size=1pt]
  coordinates {\input{Total-Latency/TotalTLSResumption1.txt} };

\addplot[line width=\abdouwidth\pgflinewidth, solid, color= red, mark size=1pt]
  coordinates {\input{Total-Latency/TotalTCPDump1.txt} };

\addplot[line width=\abdouwidth\pgflinewidth, dashed, color= magenta, mark size=1pt]
  coordinates {\input{Total-Latency/TotalTLSDump1.txt} };

\draw[->] (axis cs:90, 0.5) -- (axis cs:82, 0.5);
\node[color=black,font=\tiny] at (axis cs: 115, 0.5)  {ss2DNS};

\legend{DNS-U, NP-ss2DNS, ss2DNS, TCP-P, Offline-DNSSEC, Live-DNSSEC, DoT-R, DNS-T, DoT}
\end{axis}
\end{tikzpicture}
\caption{Total DNS Resolution}\label{fig:TotalLatency}
\end{subfigure}%
\begin{subfigure}[t]{0.27\textwidth}
    \hspace*{-0.5cm}
    \begin{tikzpicture}[scale=1]
        \begin{axis}[width=\textwidth,
            y label style={at={(axis description cs:0.04,0.5)}},
            ymajorticks=true,
            xlabel={Time (ms)},
            x label style={at={(axis description cs:0.5,0)}},
            xmin=0, xmax=1.5,
            ymin=0, ymax=1,
            ytick={0,0.2,0.4,0.6,0.8,1},
            legend pos=south east,
            legend style={font=\small},
            line width=0.4pt,
]

\addplot[line width=\abdouwidth\pgflinewidth, dashed, color= magenta, mark size=1pt]
  coordinates {\input{Processing-Latency/SPTLSDump1.txt} };

\addplot[line width=\abdouwidth\pgflinewidth, solid, color= green, mark size=1pt]
  coordinates {\input{Processing-Latency/SPUDPDump1.txt} };

\addplot[line width=\abdouwidth\pgflinewidth, dashed, color= blue, mark size=1pt]
  coordinates {\input{Processing-Latency/SPPreSECDump1.txt} };

\addplot[line width=\abdouwidth\pgflinewidth, solid, color= red, mark size=1pt]
  coordinates {\input{Processing-Latency/SPTCPDump1.txt} };
  
\addplot[line width=\abdouwidth\pgflinewidth, dashdotdotted, color= cyan, mark size=1pt]
  coordinates {\input{Processing-Latency/SPLiveSECDump1.txt} };

\addplot[line width=\abdouwidth\pgflinewidth, densely dashdotted, color= violet, mark size=1pt]
  coordinates {\input{Processing-Latency/SPsecPlusNoPriv1.txt} };

\addplot[line width=\abdouwidth\pgflinewidth, dashdotted, color= orange, mark size=1pt]
  coordinates {\input{Processing-Latency/SPsecPlus1.txt} };

\draw[->] (axis cs:0.85, 0.5) -- (axis cs:0.755, 0.5);
\node[color=black,font=\tiny] at (axis cs: 1.05, 0.5)  {ss2DNS};

\end{axis}
\end{tikzpicture}
\captionsetup{margin={0.0cm,1cm}}
\caption{Server-side processing}\label{fig:ProcessingLatency}
\end{subfigure}%

\begin{subfigure}[t]{0.27\textwidth}
\hspace*{-1.5cm}
\begin{tikzpicture}
        \begin{axis}[width=8cm,
            height = 4cm,
            y label style={at={(axis description cs:0.07,0.5)}},
            ymajorticks=true,
            xlabel={CPU Utilization (\%)},
            x label style={at={(axis description cs:0.5,0)}},
            ylabel={CDF},
            xmin=0, xmax=20,
            ymin=0, ymax=1,
            ytick={0,0.2,0.4,0.6,0.8,1},
            legend pos=south east,
            legend style={font=\normalsize},
            line width=0.4pt,
]

\addplot[line width=\abdouwidth\pgflinewidth, solid, color= green, mark size=1pt]
  coordinates {\input{CPU-util/UDP1sec.txt} };

\addplot[line width=\abdouwidth\pgflinewidth, solid, color= red, mark size=1pt]
  coordinates {\input{CPU-util/TCPutilcpu1sec.txt} };

\addplot[line width=\abdouwidth\pgflinewidth, dashed, color= blue, mark size=1pt]
  coordinates {\input{CPU-util/PreDNSSECutil2.txt} };

\addplot[line width=\abdouwidth\pgflinewidth, dashdotdotted, color= cyan, mark size=1pt]
  coordinates {\input{CPU-util/LiveDNSSEC1sec.txt} };

\addplot[line width=\abdouwidth\pgflinewidth, dashed, color= magenta, mark size=1pt]
  coordinates {\input{CPU-util/TLSutil1sec.txt} };

\addplot[line width=\abdouwidth\pgflinewidth, dashdotted, color= orange, mark size=1pt]
  coordinates {\input{CPU-util/SecPlusutil.txt} };

\addplot[line width=\abdouwidth\pgflinewidth, densely dashdotted, color= violet, mark size=1pt]
  coordinates {\input{CPU-util/NoPrivutil1.txt} };

\draw[->] (axis cs:5.21, 0.5) -- (axis cs:4.47, 0.5);
\node[color=black,font=\tiny] at (axis cs: 6.41, 0.5)  {ss2DNS};

\end{axis}

\end{tikzpicture}
\captionsetup{margin={0.0cm,-0.8cm}}
\caption{CPU utilization (120 QPS)}\label{fig:cpuutilization}
\end{subfigure}%

\caption{Total DNS resolution time, server-side processing delay, and CPU utilization of different schemes. UDP-based DNS (DNS-U), TCP-based DNS (DNS-T), TCP-P (Persistent), DoT-R (Resumption), NP-ss2DNS (No privacy). Figures b and c do not include TCP-P and \mbox{DoT-R} (see inline for explanation).}
\end{figure}

\subsection{Performance Evaluation}
\label{sec:performance}
In this section, we evaluate our prototype implementation against several other DNS schemes, focusing on three performance metrics: server-side processing time (and how it is affected with various Link-Layer MTUs), total resolution time, and CPU utilization. We setup the study using two VMs deployed on cloud servers. The aim of this evaluation is to get a preliminary assessment of how ss2DNS compares in performance to several major existing schemes in practice.

\textbf{Server-side Processing}: We used two cloud servers, each equipped with a 2-core (2.4 GHz) CPU and 2GB of RAM, running Ubuntu 20.04. Figure~\ref{fig:ProcessingLatency} illustrates the CDFs of server-side processing times for five DNS schemes compared to ss2DNS in both privacy modes. The processing time is measured as the interval between the arrival of the last DNS query datagram fragment at the ANS and the departure of the first response fragment, captured at the Link layer. While server-side processing latencies include some noise from process switching, this was considered negligible as it mainly affects the upper end of the CDFs in all the analyzed schemes.

The results of 1,000 queries for each scheme show that DNSSEC with live signing requires over 65\% more server-side processing time than UDP-based DNS at the 80th percentile. Despite the cryptographic overhead, ss2DNS processes over 90\% of queries in under one millisecond, adding approximately half a millisecond to the server-side processing time compared to UDP-based DNS. In no-privacy mode, the absence of query decryption reduces server-side processing time by $\sim$0.1 ms. Performance of ss2DNS can be further improved by reusing ephemeral keys for multiple queries within a short time, with minimal impact on forward secrecy. Aside from Live DNSSEC and ss2DNS, all other schemes exhibit similar server-side processing latency. Moreover, server-side processing times for TCP with persistent connections (TCP-P) and TLS with session resumption (DoT-R) are expected to be equivalent to standard TCP (DNS-T) and DoT, as they measure query processing time after session establishments. Thus, they are excluded from Fig.~\ref{fig:ProcessingLatency}.

\textbf{Effect of Network-Layer Fragmentation}:
Because ss2DNS messages are longer than those of Vanilla DNS and possibly DNSSEC, we examined the effect of network-layer fragmentation, particularly for messages exceeding the Maximum Transmission Unit (MTU), on the server-side processing time in all three schemes. We used four MTU values: 1450, 1000, 500, and 200.

For larger MTUs (Figs.~\ref{fig:1450MTU}, \ref{fig:1000MTU}), where fragmentation is less likely, the CDFs of Vanilla DNS and DNSSEC largely overlap, and ss2DNS experiences no fragmentation. For 200B MTU (Fig.~\ref{fig:200MTU}), both DNSSEC and Vanilla DNS are prone to fragmentation, showing similar patterns, and ss2DNS sees a $\sim$0.03 ms increase at the 90th percentile. However, at 500B (Fig.~\ref{fig:500MTU}), Vanilla DNS experiences less fragmentation, while DNSSEC remains mostly fragmented, resulting in slightly slower performance for DNSSEC. Overall, fragmentation's impact on server-side processing time is negligible.

\textbf{Total DNS Resolution Latency}: Total DNS resolution latency was measured between the two servers with an average round-trip time (RTT) of 81 ms, based on 35,000 RTT measurements, with a standard deviation of 0.131 ms. As shown in Fig.~\ref{fig:TotalLatency}, UDP-based DNS, DNSSEC, and DNSSEC with live signing have similar total latencies of approximately 82 ms. Since most ss2DNS queries are processed in under 1 ms on the server-side, the total resolution latency for ss2DNS is similar to UDP-based DNS and DNSSEC, with a negligible increase. Based on the ss2DNS measurements obtained within a single zone, it can be asserted that the overhead associated with resolving records through the DNS hierarchy is also minimal. Thus, ss2DNS has negligible impact on end-user DNS resolution experience, as the added latency is minimal. Due to the TCP handshake, TCP-based DNS results in double the latency of single round-trip schemes, while DoT is four times slower due to the extra overhead of the TLS handshake. The TLS termination messages (FIN, ACK, RST) are excluded from this calculation. If termination packets were included, DoT would be five times slower than UDP-based DNS. Although in-code timestamping for TCP-P and DoT-R was not included to measure server-side processing (Fig.~\ref{fig:ProcessingLatency}) and utilization (Fig.~\ref{fig:cpuutilization}), we measured the resolution time for these schemes (Fig.~\ref{fig:TotalLatency}) by using the timestamps of query packet departures and response packet arrivals at the resolver.

\textbf{CPU Utilization}: Figure~\ref{fig:cpuutilization} presents the CPU utilization of different DNS schemes when a multi-threaded program sends 20,000 queries at a rate of $\sim$120 queries per second. At the 90th percentile, UDP-based DNS and offline-DNSSEC exhibit identical CPU utilization, while TCP-based DNS demonstrates approximately 0.5\% higher utilization. Live DNSSEC shows an increase of 1\% in CPU usage compared to UDP-based DNS at this percentile. ss2DNS in both modes consumes approximately 2.5\% more CPU. Notably, DoT without session resumption requires around five times more CPU than UDP-based DNS and three times more than ss2DNS in both privacy modes at the 90th percentile.

\begin{figure}[t!]
\centering
\begin{subfigure}[t]{0.27\textwidth}
\begin{tikzpicture}[scale=1]
\begin{axis}[legend style={nodes={scale=0.75}},
    width=\textwidth,
    y label style={at={(axis description cs:0.15,0.5)}},
    xlabel={Server Delay (ms)},
    x label style={at={(axis description cs:0.5,0.18)}},
    xtick={0, 1.5},
    ylabel={CDF},
    xmin=0, xmax=1.5,
    ymin=0, ymax=1,
    ytick={0,0.2,0.4,0.6,0.8,1},
    legend style={at={(axis cs:1.48,0.02)},anchor=south east, fill=none},
    legend style={font=\small},
]

\addplot[line width=2\pgflinewidth, solid, color= myTeal]
  coordinates {\input{data/CDF1450plus.txt} };

\addplot[line width=2\pgflinewidth, dashed, color= myOrange]
  coordinates {\input{data/CDF1450Vanilla.txt} };

\addplot[line width=2\pgflinewidth, dotted, color= myGreen]
  coordinates {\input{data/CDF1450SEC.txt} };
  
\legend{ss2DNS, Vanilla DNS, DNSSEC} 
\end{axis}
\end{tikzpicture}

\caption{MTU = 1450B}\label{fig:1450MTU}
\end{subfigure}%
\begin{subfigure}[t]{0.27\textwidth}
\begin{tikzpicture}[scale=1]
\begin{axis}[width=\textwidth,
    y label style={at={(axis description cs:0.04,0.5)}},
    ymajorticks=false,
    xlabel={Server Delay (ms)},
    x label style={at={(axis description cs:0.5,0.18)}},
    xtick={0, 1.5},
    xmin=0, xmax=1.5,
    ymin=0, ymax=1,
    legend pos=south east,
    legend style={font=\small},
]

\addplot[line width=2\pgflinewidth, solid, color= myTeal]
  coordinates {\input{data/CDF1000plus.txt} };

\addplot[line width=2\pgflinewidth, dashed, color= myOrange]
  coordinates {\input{data/CDF1000Vanilla.txt} };

\addplot[line width=2\pgflinewidth, dotted, color= myGreen]
  coordinates {\input{data/CDF1000SEC.txt} };

\end{axis}
\end{tikzpicture}
\captionsetup{margin={0.0cm,1cm}}
\caption{MTU = 1000B}\label{fig:1000MTU}
\end{subfigure}%

\begin{subfigure}[t]{0.27\textwidth}
\begin{tikzpicture}[scale=1]
\begin{axis}[width=\textwidth,
    y label style={at={(axis description cs:0.15,0.5)}},
    xlabel={Server Delay (ms)},
    x label style={at={(axis description cs:0.5,0.18)}},
    xtick={0, 1.5},
    ylabel={CDF},
    xmin=0, xmax=1.5,
    ymin=0, ymax=1,
    ytick={0,0.2,0.4,0.6,0.8,1},
    legend pos=south east,
    legend style={font=\small},
]

\addplot[line width=2\pgflinewidth, solid, color= myTeal]
  coordinates {\input{data/CDF500plus.txt} };
 
\addplot[line width=2\pgflinewidth, dashed, color= myOrange]
  coordinates {\input{data/CDF500Vanilla.txt} };

\addplot[line width=2\pgflinewidth, dotted, color= myGreen]
  coordinates {\input{data/CDF500SEC.txt} };

\end{axis}
\end{tikzpicture}
\caption{MTU = 500B} \label{fig:500MTU}
\end{subfigure}%
\begin{subfigure}[t]{0.27\textwidth}
\begin{tikzpicture}[scale=1]
\begin{axis}[width=\textwidth,
    y label style={at={(axis description cs:0.04,0.5)}},
    ymajorticks=false,
    xlabel={Server Delay (ms)},
    x label style={at={(axis description cs:0.5,0.18)}},
    xtick={0, 1.5},
    xmin=0, xmax=1.5,
    ymin=0, ymax=1,
    ytick={0,0.2,0.4,0.6,0.8,1},
    legend pos=south east,
    legend style={font=\small},
]

\addplot[line width=2\pgflinewidth, solid, color= myTeal]
  coordinates {\input{data/CDF200plus.txt} };

\addplot[line width=2\pgflinewidth, dashed, color= myOrange]
  coordinates {\input{data/CDF200Vanilla.txt} };

\addplot[line width=2\pgflinewidth, dotted, color= myGreen]
  coordinates {\input{data/CDF200SEC.txt} };

\end{axis}
\end{tikzpicture}
\captionsetup{margin={0.0cm,1cm}}
\caption{MTU = 200B} \label{fig:200MTU}
\end{subfigure}%
\caption{Server-side processing delay for ss2DNS, Vanilla DNS, and DNSSEC in various MTUs.}
\label{fig:MTUs}
\end{figure}

\textbf{Takeaways}: Although the server-side processing latency of ss2DNS is relatively higher compared to other less secure DNS schemes, it remains $<1$  millisecond for the majority of queries. The total resolution time of ss2DNS is comparable to that of less secure schemes, as the server-side processing latency is overshadowed by network delays in the total DNS resolution time~\cite{CFRTT, crovella2006internet}. Moreover, in cases of fragmentation of large responses in ss2DNS, we showed that fragmentation has a negligible effect on server-side processing time. Finally, ss2DNS demonstrates CPU utilization comparable to that of less secure schemes, and significantly lower than DoT.

\section{Discussion}
\label{sec:discussion}
\textbf{Targeting Stage~2}: ss2DNS operates in Stage~2, requiring a secure scheme (\eg DoT~\cite{DoT-rfc7858}) in Stage~1 to secure the entire DNS resolution path. The focus on Stage~2, rather than developing a new protocol for the full resolution path, is justified by the deployment challenges faced by schemes that require fundamental changes to the original DNS design. Furthermore, various secure DNS schemes for Stage~1 have already been introduced and are increasingly adopted on both the client side (e.g., web browsers) and recursive resolver side~\cite{lu2019end, kosek2022one}. By integrating ss2DNS in Stage~2 with an existing secure Stage~1 protocol, full-path security is achieved.

\textbf{Availability of Key Servers}: Availability of key servers within each zone is critical. If a key server becomes unavailable when the short-term signing keys of nameservers expire, name resolution will fail. Since key servers play such a critical role, aside from their security, their availability also needs to be ensured through means such as server redundancy. In practice, external trustworthy key servers can be introduced in the trust model, allowing zone owners to outsource the key server role.

\textbf{Reduce Response Size}: To authenticate the ephemeral keys of the ss2DNS nameservers, these keys are signed by the short-term signing key of the nameservers. This signature is appended as part of response and used by the recursive resolvers for verifying the authenticity of ephemeral key. An alternative is to use implicitly authenticated key agreement protocols, such as MQV~\cite{mqv}. In this method, the key agreement function establishes a shared master key based on the short-term key of the nameserver ($\omega^i_l$) and the ephemeral key of the nameserver ($\Lambda^i_l$), which is implicitly authenticated. Therefore, by employing an implicit, unilateral authenticated key agreement function (\ie where only the server-side is authenticated), inclusion of the ephemeral keys' signature in the responses becomes unnecessary. As a result, the constant additional variables in ss2DNS responses can be reduced by $~$70 bytes, which further alleviates the amplification factor.

\textbf{Notifying Clients}: In the current design and implementation of ss2DNS, no means have been defined to inform clients regarding successful use of ss2DNS in Stage~2. Similar to the AD flag in DNSSEC, a DNS header bit can be defined for ss2DNS by which the clients can be informed regarding effective implementation and use of ss2DNS in Stage~2. Thereby, if a recursive resolver is trusted by a client and Stage~1 is secured, a securely-communicated confirmation to the use of ss2DNS provides the client assurance that the name resolution process completed securely.

\textbf{Mitigating Query Flooding}: Since ss2DNS is a UDP-based scheme without source IP address validation, nameservers are susceptible to query flooding, exhausting computational resources. Such attacks can be mitigated by rate-limiting techniques, forcing TCP use, or application-layer source IP address validation. Additionally, zone owners can use CDN instances for their nameservers, enabling reliable distribution of queries among nameservers.

\textbf{Delegation in the Internet}: In ss2DNS, the short-term delegation of ANSes within a zone is analogous to Delegated Credentials~\cite{DCrfc9345}. These short-term delegations mechanisms are useful in situations where a long-term secret owner does not trust all the servers hosting its service, and helps minimize attacks on the long-term secret. Short-term delegations also minimize the threat of key compromise, as they are implicitly revoked in short intervals, rendering them useless to the attackers after their expiry~\cite{AbdouRevocation20}.

\textbf{Downgrade Attacks}: Similar to downgrade attacks on HTTPS, where an attacker forces a fallback to HTTP, and mitigations are implemented outside the TLS protocol (\eg HSTS), downgrade attacks on ss2DNS to Vanilla DNS require mitigations outside of the ss2DNS protocol itself. We do not discuss this here as it is out of our current scope. However, ss2DNS is designed to fail closed (Sec.~\ref{ssec:properties}), thus mitigating \emph{within-protocol} downgrade attacks~\cite{DNSSECAgility}.

\textbf{Practical Adoption of ss2DNS}: Recent work has demonstrated the effectiveness and security of short-term delegations in CDN contexts for TLS and DTLS connections~\cite{AbdouRevocation20, DCrfc9345}. ss2DNS represents the first effort to introduce short-term delegations within the DNS context to enable real-time security. Our next step would be starting conversations with standardization bodies, including the IETF, its community, and working groups, for an initial Internet Draft.

\textbf{Facilitating Deployability}:
New security protocols often face resistance before widescale adoption, especially when confronted by the inertia of older (and often less secure) entrenched schemes, making it difficult to achieve widespread adoption. They generally require endorsement, support, or incentives from key stakeholders (such as Google spearheading the adoption of Certificate Transparency~\cite{stark2019does}) to gain traction. A more closely related example is the state of DNSSEC adoption in 2016, which was approximately 1\% among ``.com", ``.net", and ``.org"~\cite{chung2017longitudinal}, whereas around 50\% of the domains under ``.nl" and ``.cz" were DNSSEC-signed.\footnote{\url{https://www.sidn.nl/en/news-and-blogs/dnssec-adoption-heavily-dependent-on-incentives-and-active-promotion}} The main reason behind the significantly higher adoption rates of DNSSEC in ``.nl" and ``.cz" was the presence of incentive schemes that reduced fees for signed domains and promoted deployment.

As reliance on third-party and centralized DNS infrastructure providers continues to grow~\cite{trufflehunter20, querydistribution}, with approximately 20 million websites utilizing Cloudflare DNS for authoritative hosting and query resolution,\footnote{\url{https://trends.builtwith.com/ns/Cloudflare-DNS}} the adoption and deployment of ss2DNS by these providers could significantly facilitate and accelerate its widespread implementation on the ANS and resolver sides. As demonstrated by the effectiveness of incentive schemes in promoting DNSSEC adoption, employing similar incentives for infrastructure providers may be a viable approach to enabling the large-scale deployment of ss2DNS.

Since DNSSEC has been adopted by the root and most TLDs~\cite{yajima2021measuring}, the ss2DNS trust model can leverage DNSSEC’s existing chain of trust for authenticated zone key retrieval, thereby enhancing deployability. Distributing ss2DNS keys as signed DNSSEC records in parent zones allows ss2DNS-enabled zones to remain compatible with DNSSEC-enabled parents, supporting incremental deployment without requiring immediate adoption by all parent zones. Additionally, automation techniques such as those outlined in RFC 7344~\cite{rfc7344} (Automating DNSSEC Delegation Trust Maintenance) have been proposed to facilitate the publication of DS records in DNSSEC. Given that ss2DNS employs a PKI and trust model similar to DNSSEC, we believe that these automation mechanisms can be used to support the deployment of ss2DNS.

\section{Related Work}
\label{sec:RelatedWork}
\textbf{Threats and mitigations in Stage~2}: Since Kaminsky~\cite{KaminskyAttack} demonstrated the weakness of resolvers to off-path cache poisoning, solutions that introduce more randomness to DNS messages, such as~\cite{rfc5452, dagon2008encoding, perdisci2009wsec}, have been proposed. Since on-path adversaries have access to the included randomness in DNS queries and responses, these randomness-based solutions can only be effective against off-path adversaries. Moreover, researchers have demonstrated attacks that lead to inferring or bypassing the random values included in DNS messages by off-path adversaries. For example, Herzberg~\etal~\cite{herzberg2012security} introduced a technique for predicting the source ports of queries of resolvers behind a Network Address Translation (NAT). In another research, Herzberg~\etal~\cite{herzberg2013fragmentation} demonstrated a method for bypassing source port randomization of responses, when the responses from ANSes are fragmented. Additionally, Man~\etal~\cite{poisoningReloaded} used network side-channels for inferring the DNS query source ports and cache poisoning.

\textbf{Other Stage~2 schemes}: Beyond adding more entropy to DNS responses to mitigate off-path cache poisoning, solutions like DNSSEC~\cite{rfc4033, rfc4034, rfc4035}, add message authentication to mitigate cache poisoning by both off- and on-path adversaries. However, DNSSEC adoption remains low~\cite{DNSSECAgility, chung2017longitudinal}, and recent studies~\cite{DNSSECAgility} show that ambiguities in its specification have led to vulnerable resolver implementations that accept unverifiable DNSSEC records, exposing them to cache poisoning via false responses. Such vulnerabilities can be exploited by attackers for false response injection and cache poisoning~\cite{DNSSECAgility}. DNSCurve~\cite{DNSCurveweb, curvedraft} another Stage~2 scheme, was also proposed but never widely adopted~\cite{IanaDNS, SchmidThirty}. RHINE is another secure DNS scheme proposed by Duan~\etal~\cite{duan2023rhine}, which relies on a hybrid trust model, where the web PKI is used but with the root zone of DNS remaining an authority by self-signing its certificate. RHINE provides authenticated zone delegation by keeping the global delegation status of DNS, and offers message authentication using pre-signed zone records. Some Stage~1 schemes, such as DoT~\cite{DoT-rfc7858} and DNS-over-QUIC~\cite{doq-rfc9250}, have been proposed to be used in Stage~2 as well. However, root and TLD authorities are reluctant to rely on third-party Certification Authorities (CAs) as part of their trust model. Confidential DNS, proposed as an Internet-Draft to enhance DNS privacy across both stages, did not advance~\cite{confdns}. It offers opportunistic encryption using new key records, but its unauthenticated form is vulnerable to active attacks, and its authenticated version depends on DNSSEC~\cite{confdns}.

\section{Concluding Remarks}
\label{sec:Conclusion}
Herein, we presented ss2DNS, a secure DNS scheme in Stage~2, which relies on a DNSSEC-like trust model. ss2DNS not only provides more robust security properties but also demonstrates a relatively similar name resolution performance compared to the previously proposed schemes. The minimal DNS resolution latency in ss2DNS is a result of considering single round-trip as one of the design properties, which was thoroughly discussed and justified within this paper. ss2DNS avoids duplicating the long-term keys on the nameservers within a zone, addressing the concern that certain zone owners (\eg root) may not trust all the nameservers that serve their zone data. Moreover, ss2DNS is compatible with Vanilla DNS, as the zone files and DNS record lookup function remain unchanged on the server side. To enhance the security of the entire DNS resolution path, it is recommended to use ss2DNS alongside one of the secure DNS schemes from Stage~1.

\bibliographystyle{IEEEtran}

\bibliography{Bibliography}

\end{document}